\documentclass[conference]{IEEEtran}
\IEEEoverridecommandlockouts
\usepackage{cite}
\usepackage{amsmath,amssymb,amsfonts}
\usepackage{algorithmic}
\usepackage{graphicx}
\usepackage{textcomp}
\usepackage{xcolor}
\usepackage{placeins}
\usepackage{url}
\usepackage[caption=false,font=normalsize,labelfont=sf,textfont=sf]{subfig}
\def\BibTeX{{\rm B\kern-.05em{\sc i\kern-.025em b}\kern-.08em
    T\kern-.1667em\lower.7ex\hbox{E}\kern-.125emX}}
\begin{document}

\title{Towards Production-Worthy Simulation for Autonomous Cyber Operations}

\author{
\IEEEauthorblockN{
Konur Tholl\IEEEauthorrefmark{1},
Fran\c{c}ois Rivest\IEEEauthorrefmark{2},
Mariam El Mezouar\IEEEauthorrefmark{2},
Adrian Taylor\IEEEauthorrefmark{3},
Ranwa Al Mallah\IEEEauthorrefmark{4}
}

\IEEEauthorblockA{\IEEEauthorrefmark{1}
Department of Electrical and Computer Engineering\\
Royal Military College of Canada\\
Kingston, Canada}

\IEEEauthorblockA{\IEEEauthorrefmark{2}
Department of Mathematics and Computer Science\\
Royal Military College of Canada\\
Kingston, Canada}

\IEEEauthorblockA{\IEEEauthorrefmark{3}
Defence Research and Development Canada\\
Ottawa, Canada}

\IEEEauthorblockA{\IEEEauthorrefmark{4}
Department of Computer and Software Engineering\\
Polytechnique Montreal\\
Montreal, Canada}
}

\maketitle

\begin{abstract}
Simulated environments have proven invaluable in Autonomous Cyber Operations (ACO) where Reinforcement Learning (RL) agents can be trained without the computational overhead of emulation. These environments must accurately represent cybersecurity scenarios while producing the necessary signals to support RL training. In this study, we present a framework where we first extend CybORG's Cage Challenge 2 environment by implementing three new actions: Patch, Isolate, and Unisolate, to better represent the capabilities available to human operators in real-world settings. We then propose a design for agent development where we modify the reward signals and the agent's feature space to enhance training performance. To validate these modifications, we train DQN and PPO agents in the updated environment. Our study demonstrates that CybORG can be extended with additional realistic functionality, while maintaining its ability to generate informative training signals for RL agents.
\end{abstract}


\section{Introduction}
Current applications in Autonomous Cyber Operations (ACO) rely on Reinforcement Learning (RL) to train agents capable of making effective decisions with minimal human oversight \cite{wiebe_learning_2023,mcdonald_competitive_2023,baillie_cyborg_2020,loevenich_designllmcyborg_2024}. To train these agents, an environment must exist that mimics the cybersecurity domain, and produces reliable signals for RL. The Technical Cooperation Program (TTCP) has created such an environment: CybORG's Cage Challenge 2 \cite{baillie_cyborg_2020,cagechallenge2github}, which is a popular RL environment designed for ACO. 

As with all simulated ACO environments, the agents trained using CybORG are not used in operational settings \cite{oesch_pathtoaco_2025}. This lack of practical adoption primarily stems from the mismatch between RL and real-life cybersecurity conditions, where defenders and adversaries are not constrained to predefined rules or a limited action space. The primary focus of this study is to take a step toward operationally relevant autonomous cyber agents by modifying CybORG's action space.
  
 In particular, we extend the environment by adding new actions that better reflect the variety of options available in operational settings. To evaluate the success of our study, we develop baseline RL agents and demonstrate that they can be efficiently trained within a more realistic cybersecurity simulation.

Specifically, the contributions this study makes are:
\begin{itemize}
    \item \textit{CybORG Augmentation with Additional Realistic Actions}. We extend the CybORG environment by adding three distinct actions: Isolate, Unisolate and Patch, each representing capabilities common to human operators. 
    \item \textit{RL Agent Implementation}. We develop and train Proximal Policy Optimization (PPO) and Deep Q-Network (DQN) agents, evaluating them across combinations of hyperparameters.
    \item \textit{RL Signal Optimizations}. We modify CybORG's feature space and reward signals to improve training efficiency and performance. 
\end{itemize}

This paper is based on the research presented in the author's Master's thesis \cite{tholl_thesis_2025}.

\section{Background}
In this section, we provide a high-level overview of related fields and analyze previous work to identify the novel contributions addressed in this study. 

\subsection{RL and Cybersecurity}
The magnitude and sophistication of offensive cyber threats have made it infeasible for people to manually defend their systems, making automated tools essential \cite{oesch_pathtoaco_2025}. 

These tools began as signature-based approaches, which involve storing known attacks and comparing them against incoming threats. The requirement for the exploit to already exist in a database for successful detection means the system will fail for zero-day vulnerabilities. Moreover, the vast, rapidly increasing quantity of possible attacks makes it difficult to maintain a repository of exact matches. 

Machine Learning (ML) has shown promise in addressing the limitations of signature-based approaches, where models can be trained to detect patterns across malicious samples, eliminating the requirement for exact matches \cite{center_for_security_and_emerging_technology_machine_2021}. However, ML typically depends on large historical datasets to train these models, which are difficult to maintain in the ever-evolving cyber domain. Moreover, the models are still susceptible to zero-day vulnerabilities, as exploits may lack recognizable patterns within the training data.

RL is its own ML paradigm that eliminates this reliance on datasets; instead, agents learn by directly interacting with their environment \cite{qiang_reinforcement_2011}. The agent executes an action on the environment, which produces a corresponding reward, and next state. The agent then uses these signals to compute a loss function to optimize its parameters. There are many algorithms for RL, with the fundamental differences being how they compute the loss using these signals; in this study, we focus on PPO and DQN \cite{schulman_proximal_2017,lyu_advance_2022}.  

Current applications of ACO leverage RL, where agents are trained in environments to learn policies for making effective decisions \cite{Faizan_MARL_2024,wiebe_learning_2023,loevenich_designllmcyborg_2024}. Agents are able to adapt to adversaries' dynamic behavior without requiring manual updates to training datasets.

\subsection{CybORG}
While RL eliminates the requirement to create and maintain datasets, it requires a suitable environment that produces the necessary signals to train agents; in this study, we use CybORG' Cage Challenge 2 - to keep this paper concise we will simply refer to it as CybORG\cite{baillie_cyborg_2020, cagechallenge2github}.

CybORG is an RL environment for cybersecurity that supports both simulation and emulation. Unlike emulation, which has agents train on physical hosts, simulation means that there is no physical replication of a system, instead objects exist with their own instance variables that represent the system state. As the scenario progresses, these variables are updated, mimicking real-life cybersecurity conditions \cite{baillie_cyborg_2020}. Simulation was ultimately used for our study due its resource efficiency, simpler implementation, and high scalability. 

In particular, the goal of CybORG is to train blue agents to defend a simulated enterprise network. The agent selects actions from a predefined action space to prevent an adversary (the red agent) from compromising the network. Each of these actions produces a reward and state, which the blue agent can use to learn an optimal policy for selecting subsequent actions. These rewards are computed every timestep based on the state of the network, and are always zero or less \cite{cagechallenge2github}. This presents an opportunity to alter these signals to enable the agent to better distinguish between good and bad actions.

Furthermore, CybORG's action space does not contain the range of actions that would be available to blue operators. It also contains six deception-based actions that, while important to cybersecurity, are not entirely relevant from an ACO perspective. Modifying this action-space could enable CybORG to better mirror real-world cybersecurity settings \cite{baillie_cyborg_2020}.

Finally, previous work typically deploys value-based or policy-based RL algorithms for CybORG without explicit evaluation of each \cite{loevenich_designllmcyborg_2024,wiebe_learning_2023,mcdonald_competitive_2023,Faizan_MARL_2024}. While A. Dutta et al. compared different RL algorithms in cybersecurity environments, they did not specifically use CybORG \cite{dutta_rlevaluationnotcyborg_2023}. This presents an opportunity to directly compare value-based and policy-based algorithms within the context of CybORG.

\section{Methodology}
In this section, we explain the design and implementation of our work. In particular, we discuss:

\begin{itemize}
    \item Our modifications to the CybORG action-space and the implementation details of the Patch, Isolate and UnIsolate actions.
    \item How we altered CybORG's feature space and reward signals to support these new actions and optimize training.
    \item The design of the baseline agents and how they were used to evaluate our work.
\end{itemize}

\subsection{Action Space Modification}
\label{meth:actionmodification}
To better reflect real-world cybersecurity, we added three actions to CybORG. These include:
\begin{itemize}
    \item Patch. Decreases the chance of an attacker's exploit succeeding on a host.
    \item Isolate. Simulates unplugging the host from the network - preventing connectivity to or from any host.
    \item UnIsolate. Simulates rejoining an isolated host back to the network.
\end{itemize}

\subsubsection{Patch Action}
We implemented the Patch action to simulate updating vulnerable services on a host, decreasing the chances of the host being exploited. Each host is assigned a patch score between 0 and 1, which represents how likely an attacker's exploit is to succeed (0 being guaranteed success). At the start of an episode, all patch scores are initialized to 0.

To enforce patching, the red agent's PrivilegeEscalate and ExpoitRemoteService were modified to have an \textit{X} percent chance of failing, where \textit{X} corresponds to the host's patch score. This randomness helps mimic the stochastic nature of cybersecurity. Moreover, these actions decrease the patch score to simulate reconnaissance and prevent blue from continually exploiting it. Specifically, red actions decrease the patch score by 0.35, whereas the Patch action increases it by 0.3.

\begin{figure}[!b]
    \centering
    \captionsetup{justification=raggedright, singlelinecheck=false} 
    \includegraphics[width=\columnwidth]{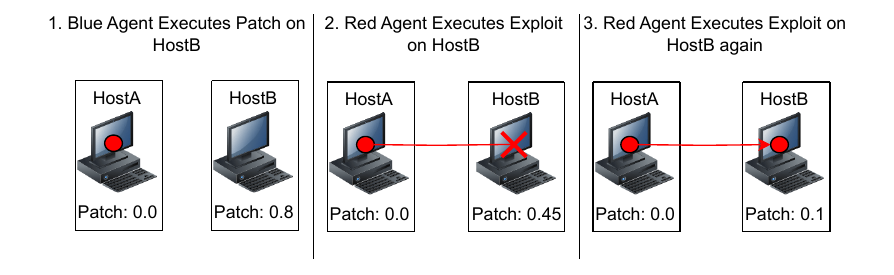}
    \caption[Design of the Patch Action.]{Diagram illustrating the functionality of the Patch action. The blue agent patches HostB (left), increasing its patch score. The red agent attempts to exploit HostB (middle), failing the exploit, but decreasing the patch score. The red agent attempts to exploit HostB again, this time successfully establishing a session.}
    \label{fig:patchActionOverview}
\end{figure}

Fig. \ref{fig:patchActionOverview} shows our implementation of the Patch action. 

\subsubsection{Isolate Action}
We implemented the Isolate action to simulate unplugging a host from the network, preventing any connectivity to or from that host.

We tracked isolated hosts using a list of bits, where the index of each bit corresponds to a host - 1 for isolated, 0 for unisolated. At the start of the episode, all hosts are unisolated. 

The red agent's actions were modified to respect isolated hosts. Specifically, DiscoverNetworkServices, ExploitRemoteService, PrivilegeEscalate, and Impact were set to fail if the source or destination was isolated, and DiscoverRemoteSystems (i.e., ping sweep) returned nothing for isolated hosts.

We applied a negative reward to simulate the network disruptions associated with isolating hosts. This was proportional to the priority of the host being isolated: -0.2 for user, -0.4 for enterprise, and -0.5 for operational (the attacker's end target).

To simulate persistence, our implementation of Isolate does not destroy any existing red sessions. As soon as a compromised host is unisolated (our action to simulate joining a host back to the network), it can be used to further red's attack; malicious artifacts must be removed via the Remove or Restore actions. 

\begin{figure}[!b]
    \centering
    \captionsetup{justification=raggedright, singlelinecheck=false} 
    \includegraphics[width=\columnwidth]{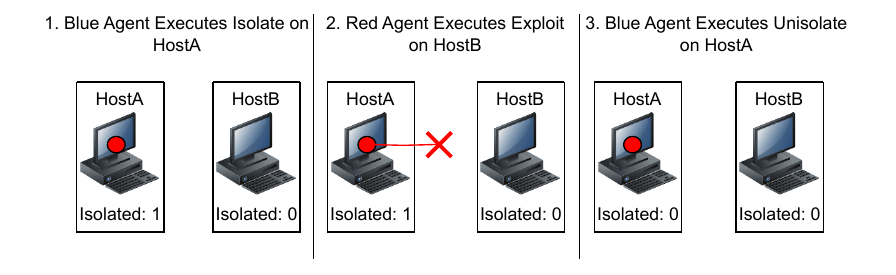}
    \caption[Design of the Isolate Action.]{Illustration of the Isolate action's functionality. The blue agent isolates HostA (left), disconnecting it from the network. The red agent attempts to exploit HostB from HostA (middle), but cannot connect since it is isolated. The blue agent unisolates HostA (right), but the red agent's session remains, as it was not explicitly removed.}
    \label{fig:isolateActionFunctionality}
\end{figure}

Fig. \ref{fig:isolateActionFunctionality} illustrates the functionality of the Isolate and Unisolate actions.

\subsubsection{Action Removal}
CybORG includes deception-based actions, which set up their own decoy services (honeypots). Honeypots are practical in real-life operations and enable blue teams to gain insights into adversaries' behaviors to fortify future defenses \cite{otal_honeypot_2024}. However, from an ACO perspective, these actions are not ideal and do not have the same impact as more proactive actions like Restore or Remove. To use honeypots effectively, they should be pre-established as a dedicated environment for training agents, instead of as available actions. As such, we removed these actions for our study.

We also removed the Monitor action from CybORG. This is because it is called after every timestep to provide the blue agent with the updated network state. This choice was motivated to avoid redundant behavior, rather than to mimic cybersecurity conditions.

Overall, the action space used for our research consists of six distinct actions: Analyze, Restore, Remove, Isolate, Unisolate and Patch. The functionality of Analyze, Restore, and Remove were unchanged from the original CybORG environment.

\subsection{Preprocessing}
\label{meth:preprocessing}
We added the hosts' isolated and patch status to the agent's state space. Isolated is represented as a 0 or 1, and Patch as the host's patch score. We also changed the existing activity and compromised state preprocessing from one-hot encoding to being represented as a normalized float. The primary reason for this is to reduce the feature space, but it also ensures that the agent will inherently prioritize certain categories due to their increased impact on the gradient. For example, if the Scan state is 0.3 and the Exploit state is 1.0, the agent will automatically prioritize the Exploit state, as its value contributes more to the agent's training and its resulting action selection. 

We also appended two additional floats to the state space representing the number of isolated hosts, and the number of compromised hosts. These were normalized by dividing by the total number of hosts, to ensure these features weren't inherently prioritized.

\begin{figure}[!b]
    \centering
    \captionsetup{justification=raggedright, singlelinecheck=false} 
    \includegraphics[width=\columnwidth]{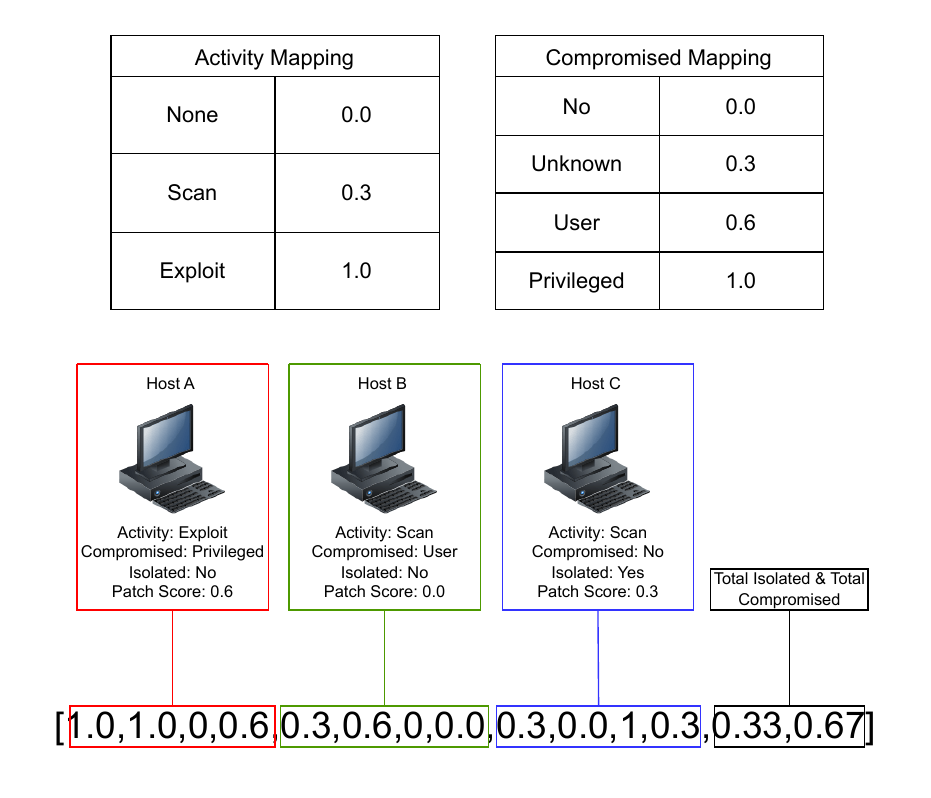}
    \caption[Modifying the Default Feature Space Mapping.]{Modified feature space mapping used in this study.}
    \label{fig:featureSpaceMappingOrigToNew}
\end{figure}

Fig. \ref{fig:featureSpaceMappingOrigToNew} shows the updated feature space mapping employed in this study.

\subsection{Signal Modification}
\label{meth:sigmodification}
We normalized CybORG's reward signal to fall between \([-2.5,2.5]\), rather than using the original implementation, where the agent receives a negative reward at every timestep. This updated signal provides a clearer distinction between desirable and undesirable actions, thereby supporting the agent's learning. Moreover, this change was designed to support future teacher-guided techniques, as a constant negative reward could push the agent's policy away from the teacher's guidance during early training.

In particular, we normalized the reward using:
\[
r_{t}^{new}=\frac{r_{t}+13.1}{13.1}*5-2.5
\]

Where \(r_{t}^{new}\) is the updated reward and \(r_{t}\) is the reward returned by CybORG. -13.1 was recorded as the highest negative reward experienced by a randomized policy across 100,000 timesteps. As such, adding 13.1 to the reward and dividing it by the same value will ensure the signal is normalized between 0 and 1. Multiplying by 5 and subtracting 2.5 ensures that this signal is transformed to fall in \([-2.5,2.5]\).

\subsection{Baseline Agent Development}
\label{meth:baselineagent}
Baseline DQN and PPO agents were developed to evaluate the performance of our modifications. Both of these agents were based heavily on existing work \cite{github_abhishek_reinforcement_learningatari_dqn_imageipynb_nodate,github_keon_deep-q-learningdqn_batchpy_nodate,lyu_advance_2022, antonin_raffin_dlr-rmstable-baselines3_2025, schulman_proximal_2017, tabor_youtube-code-repositoryreinforcementlearningpolicygradientppotorch_2021}. The fundamental difference between these two algorithms lies in how actions are selected and how the loss is computed. 

\subsubsection{DQN}
For DQN, the agent outputs a value indicating how favorable every action is for a particular state, then greedily selects the action with the highest value (Q-value). We used standard epsilon-greedy to strike a balance between exploration and exploitation \cite{lyu_advance_2022}. 

Like most DQN implementations, the loss was computed as the Mean Square Error (MSE) between the agent's current prediction and the estimated returns \cite{github_keon_deep-q-learningdqn_batchpy_nodate, github_abhishek_reinforcement_learningatari_dqn_imageipynb_nodate}. In particular, the loss was computed as \cite{sharma_dqn_2023}: 
\[
\mathbb{E}[(r+\gamma \max\limits_{a'}Q(s';a';\theta') -Q(s,a;\theta))^2]
\]

where \(r\) is the reward for selecting action \(a\), \(\theta\) represents the network's parameters, \(\gamma\) is the discount factor that ensures future actions are weighed less, and \(\max\limits_{a'}Q(s';a';\theta')\) is the maximum Q-value of the next state. A different MLP, \(\theta'\) was used to compute the target. This was updated less frequently than the primary MLP to facilitate smoother updates (i.e., to avoid chasing a moving target). 

\begin{figure}[!t]
    \centering
    \captionsetup{justification=raggedright, singlelinecheck=false} 
    \includegraphics[width=\columnwidth]{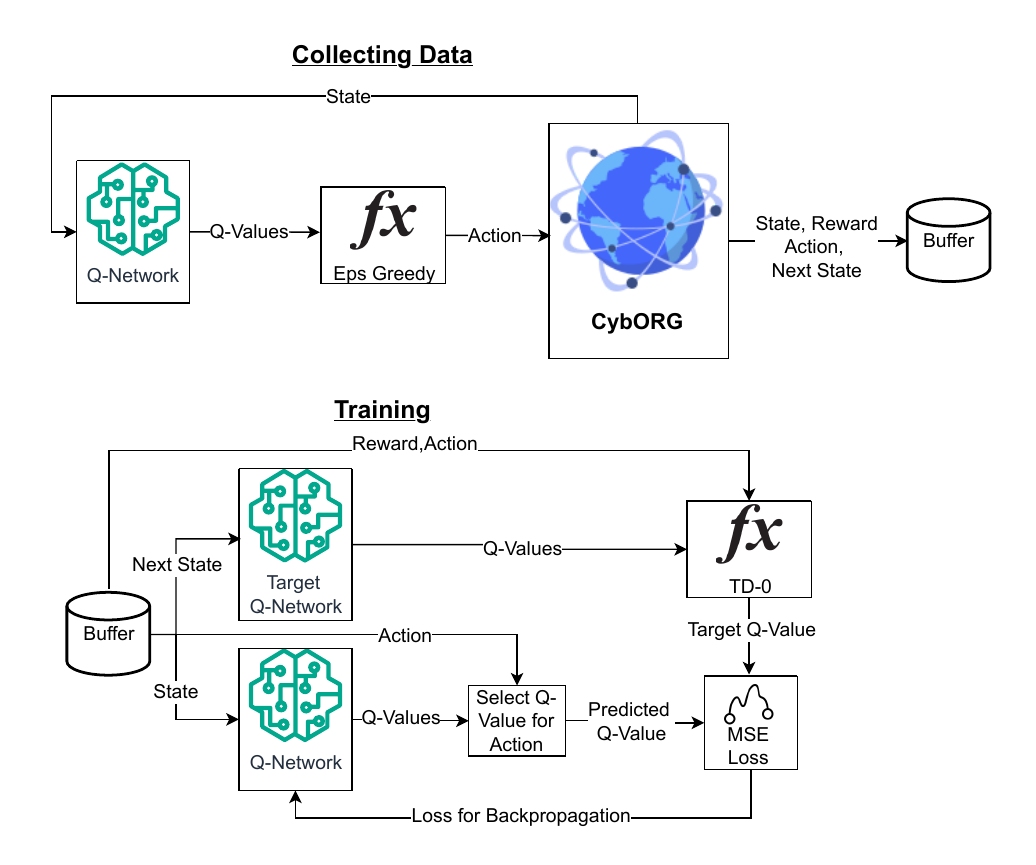}
    \caption[DQN Design.]{The DQN implementation used for this research. The data collection process is shown at the top, where the agent interacts with the environment to gather training samples. The training process is shown at the bottom, where the agent uses the collected data to compute the MSE loss between its prediction and the actual returns. In reality, the training is done in batches; however, this is omitted from the diagram for readability.}
    \label{fig:DQNImplementation}
\end{figure}

Fig. \ref{fig:DQNImplementation} illustrates the DQN implementation used in our study.

\subsubsection{PPO}
Unlike DQN, PPO outputs a probability distribution across actions, and then directly samples from this distribution. The stochastic nature of probabilities eliminates the requirement to incorporate additional exploratory techniques.

We used PPO's clipped-surrogate loss implementation, to ensure updates were kept stable \cite{schulman_proximal_2017}:

\[
L_{A}(\theta)=\mathbb{E}[min(r_{t}(\theta)A_{t},clip(r_{t}(\theta),1-\epsilon,1+\epsilon)A_{t})]
\]

where \(\epsilon\) is the policy clipping value and \(A_{t}\) represents the advantages. For computing \(A_{t}\), we used Stable Baseline 3's version of the Generalized Advantage Estimate (GAE), which strikes a balance between Monte Carlo (MC) and Temporal Difference (TD) for solving the Bellman's equation to account for future rewards \cite{antonin_raffin_dlr-rmstable-baselines3_2025}.

PPO employs a separate critic network to compute these advantages, which is trained using the same MSE loss as DQN:

\[
L_{c}(\phi)=\mathbb{E}[((V_{old}(s_{t})+A_{t})-V_{new}(s_{t}))^2]
\]

where \(V_{old}(s_{t})\) are the old critic values, and \(V_{new}(s_{t})\) are the new critic values.

\begin{figure}[!b]
    \centering
    \captionsetup{justification=raggedright, singlelinecheck=false} 
    \includegraphics[width=\columnwidth]{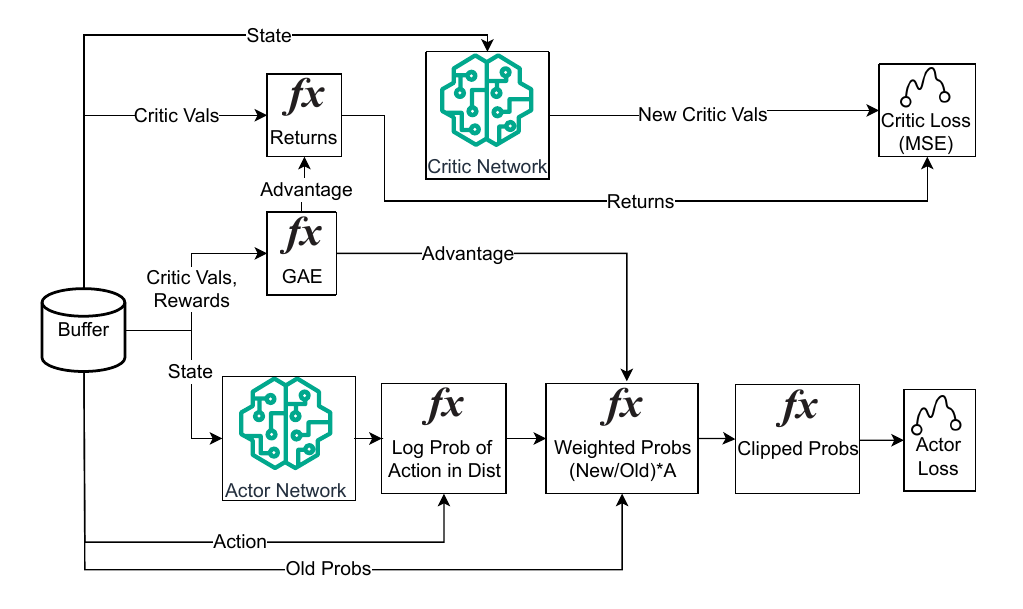}
    \caption[PPO Design - Training]{The training process for the PPO implementation. This illustrates how the critic loss and actor loss are computed using the sampled data.}
    \label{fig:ppoTrainingOverview}
\end{figure}

We present the training process for this study's implementation of PPO in Fig. \ref{fig:ppoTrainingOverview}.

\section{Evaluation}
In this section, we explain the rationale behind implementation decisions and present the results of our work.

\subsection{Choosing the RL Algorithm}
To evaluate the PPO and DQN implementations discussed in Section \ref{meth:baselineagent}, we performed a grid search over many combinations of hyperparameters. In particular, PPO was evaluated over 162 unique combinations of batch size (8, 16), training interval (16, 32, 64), actor and critic network learning rates ($e^{-4}$, $5e^{-4}$, $e^{-3}$), and policy clip (0.1, 0.15, 0.2). DQN was evaluated over 486 combinations of batch size (8, 16), queue size (100, 200, 300), learning rate ($e^{-4}$, $5e^{-4}$, $e^{-3}$), discount factor (0.85, 0.9, 0.95), epsilon (0.9, 0.95, 0.99), and epsilon decay (0.98, 0.99, 0.995).

The reason that DQN had approximately three times as many configurations as PPO is that it involved a larger number of hyperparameters. As a result, more experiments were required to cover the search space. The mean reward over the last 10 episodes, averaged across five runs was used to determine the best-performing model. Each run consisted of 150 episodes.

Tables \ref{tab:dqnresults} and \ref{tab:pporesults} presents the evaluation results for DQN and PPO. As shown, PPO consistently outperformed DQN, despite having half the training runs. 

\begin{figure}[!b]
    \centering
    \captionsetup{justification=raggedright, singlelinecheck=false} 
    \includegraphics[width=\columnwidth]{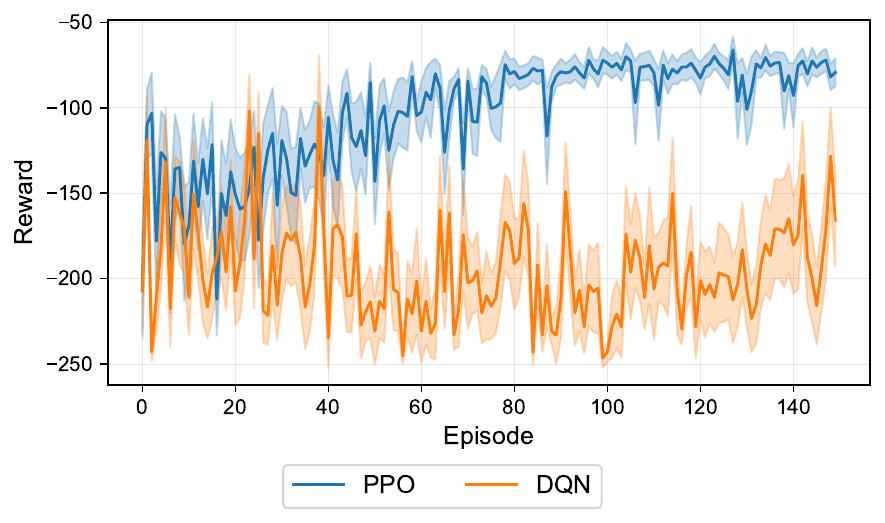}
    \caption{Comparing the top-performing PPO and DQN configurations.}
    \label{fig:ppovsdqn}
\end{figure}

To further our analysis, we saved the plots for each of these combinations and observed that PPO achieves smoother convergence. We present the plots of the top combination for each algorithm in Fig. \ref{fig:ppovsdqn}. The initial randomness exhibited by the DQN agent is expected due to the nature epsilon-greedy exploration. However, there is a 4.69\% probability of selecting a random action by episode 150, indicating that the lack of convergence is not solely attributable to the epsilon-greedy implementation.

\begin{table}[!t]
    \centering
    \caption{DQN performance: top 3 results are shown, ordered by their mean reward. Titles are abbreviated for conciseness.}
    \begin{tabular}{|c|c|c|c|c|}
        \hline
        \textbf{Batch/Queue} & \textbf{LR/Discount} & \textbf{Epsilon/Decay} & \textbf{Reward} \\
        \hline
        8/300 & 0.0010/0.90 & 0.95/0.995  & -107.74 \\
        8/200 & 0.0010/0.90 & 0.95/0.995 & -136.00 \\
        8/300 & 0.0010/0.85 & 0.95/0.99  & -141.28 \\
        \hline
    \end{tabular}
    \label{tab:dqnresults}
\end{table}

\begin{table}[!b]
    \centering
    \caption{PPO performance: top 3 results are shown, ordered by their mean reward. Titles are abbreviated for conciseness.}
    \begin{tabular}{|c|c|c|c|c|}
        \hline
        \textbf{Batch/Train} & \textbf{LR (C/A)} & \textbf{Clip} & \textbf{Reward} \\
        \hline
        16/16 & 0.0010 /0.0010 & 0.15 & -48.484 \\
        8/16 & 0.0005/0.0010 & 0.10 & -50.278 \\
        8/32 & 0.0010/0.0005 & 0.15 & -53.390 \\
        \hline
    \end{tabular}
    \label{tab:pporesults}
\end{table}

\subsection{Optimizing PPO}
Fig. \ref{fig:ppovsdqn} shows the quick convergence of our PPO implementation, but it appears to plateau at roughly episode 100. Furthermore, it plateaus at different values across runs, which is indicative of convergence to local minima. To address these issues, we repeated hyperparameter tuning using Optuna, an open-source Bayesian optimization framework \cite{noauthor_optuna_nodate}. We also added an entropy bonus to the loss and applied gradient clippings to support stable updates \cite{antonin_raffin_dlr-rmstable-baselines3_2025}. Table \ref{tab:ppooldvsnew} lists the differences between the updated and original PPO hyperparameters.

\begin{table}[!t]
\centering
\caption{Comparison of hyperparameters between the original and updated PPO implementations. A dash '-' indicates that the hyperparameter was not applied}
\small
\begin{tabular}{|l|c|c|}
\hline
\textbf{Hyperparameter}     & \textbf{Old PPO Agent} & \textbf{New PPO Agent} \\
\hline
Episode Size                & 32    & 32 \\
\hline
Batch Size                 & 16     & 256    \\
\hline
Training Interval Size         & 16     & 256    \\
\hline
Critic LR                  & 0.001  & 0.0016 \\
\hline
Policy LR                  & 0.001  & 0.0016 \\
\hline
Epochs                     & 30     & 30     \\
\hline
Policy Clip                & 0.15   & 0.2    \\
\hline
Entropy Coef               & 0      & 0.005  \\
\hline
Entropy Decay              & 0      & 0.99   \\
\hline
Critic Grad Clip           & -   & 0.1    \\
\hline
Policy Grad Clip           & -    & 0.5    \\
\hline
\end{tabular}
\label{tab:ppooldvsnew}
\end{table}

\begin{figure}[!b]
    \centering
    \captionsetup{justification=raggedright, singlelinecheck=false} 
    {\scriptsize\textbf{(a)} Original Environment}\par
    \includegraphics[width=1.0\columnwidth]{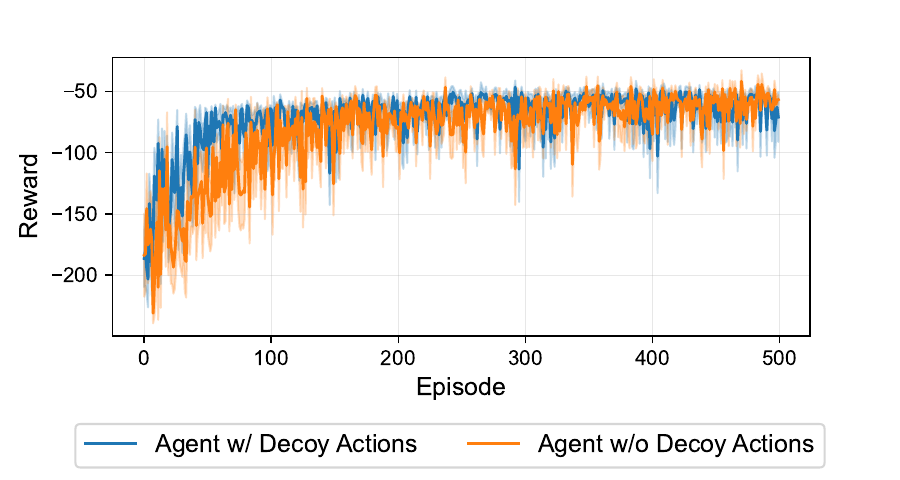}\par
    \vspace{1ex}

    {\scriptsize\textbf{(b)} Modified Environment}\par
    \includegraphics[width=1.0\columnwidth]{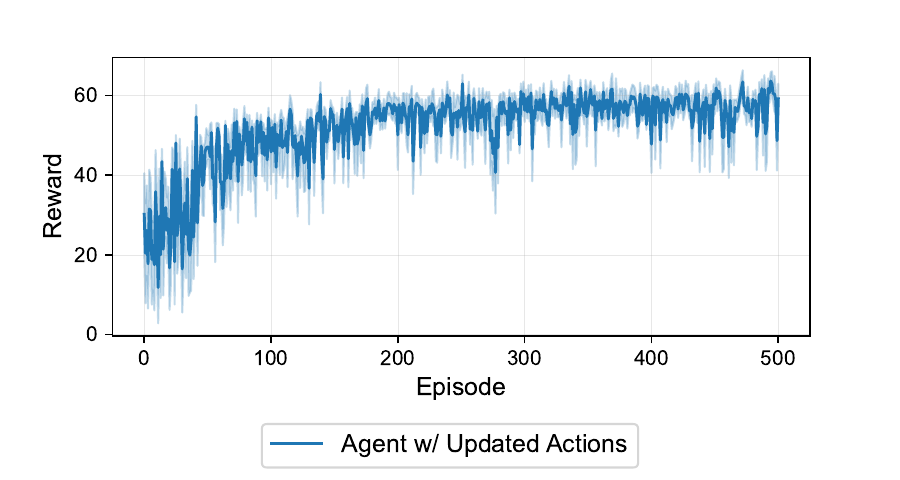}

    \caption{PPO training in original environment (top) versus the modified environment (bottom). Mean reward over 10 independent runs with a +/- 1 SE.}
    \label{fig:OrigEnvvsNewEnv}
\end{figure}

\subsection{Environment Modification}
To confirm whether our modifications described in Section \ref{meth:actionmodification} still enable efficient learning, we trained a PPO agent in the original and updated environment, as shown in Fig. \ref{fig:OrigEnvvsNewEnv}.

We can see that in both environments, the agent is able to converge onto an approximately equivalent favorable policy, demonstrating the potential to create more realistic simulations while providing effective RL signals. It should be noted that with our reward normalization, the performance of the updated environment is at a smaller scale. For example, an increase of -200 to -50 in the original environment is equivalent to an increase of 3.68 to 60.92 in the new environment.

\section{Conclusion}
ACO is an innovative field aimed at training cyber agents capable of autonomously executing actions on behalf of humans; however, these agents are not used in operational settings \cite{oesch_pathtoaco_2025}. This paper takes a step toward more realistic simulated environments by extending CybORG's action space and modifying its state space and reward signals to optimize training.

\subsubsection{Contributions}
Our primary contribution is the extension of the CybORG environment with additional blue actions to enable more representative training for cybersecurity. In addition, we provide (i) an empirical evaluation showing that PPO outperforms DQN for ACO, and (ii) modifications to CybORG's reward signals and state space to improve training efficiency.

\subsubsection{Limitations}
While we made contributions to ACO in this study, there are several limitations that should be noted.

\textit{Agent Evaluation.}
While we evaluated agent performance across many combinations of hyperparameters, the 150-episode runs may have ended before the agent fully converged to a policy, potentially decreasing the overall quality of the evaluation.

\textit{State Representation.}
Our modifications to the process for converting CybORG's raw output to numerical features were minimal and still involved computing compromised and activity states for each host using basic conditional statements, potentially underrepresenting and oversimplifying the complex state space of cybersecurity.

\textit{Unrealistic Environment.}
While our primary contribution was adding actions to better represent the vast options available to operators, the environment still has adversaries and defenders execute actions in an unrealistic sequential fashion, which under-represents real-world network activity, and simplifies adversarial attack patterns.

\subsubsection{Future Work}
Even though our work is a step toward more realistic ACO environments, there is significantly more progress required before simulated environments can sufficiently train operational agents. Several opportunities exist to help bridge this gap, including:
\begin{itemize}
    \item Increasing the number of processes and files on hosts to better mimic typical system activity.
    \item Breaking the sequential fashion in which red and blue agents execute actions - for example, by having them act until an end condition is met.
    \item Improving the mapping between CybORG's output and the RL agent's features - potentially by incorporating a neural network to capture non-linear patterns.
\end{itemize}

Taken together, these opportunities highlight ways in which RL environments for ACO can be improved to produce production-quality agents, and our work represents an initial step toward realizing this goal.

\newpage
\bibliographystyle{unsrt}
\bibliography{skeleton}

\end{document}